# Exploiting ITO colloidal nanocrystals for ultrafast pulse generation

Qiangbing Guo[†,‡], Yudong Cui[‡,§], Yunhua Yao[#], Yuting Ye[†,‡], Yue Yang[†,‡], Xueming Liu[‡,§],

Shian Zhan[#], Xiaofeng Liu[†,‡,*], Jianrong Qiu[‡,§,*] and Hideo Hosono[II]

[†]Institute of Inorganic Materials, School of Materials Science & Engineering, Zhejiang University, Hangzhou 310027, China.

[‡]State Key Laboratory of Modern Optical Instrumentation, School of Materials Science & Engineering, Zhejiang University, Hangzhou 310027, China.

[§]College of Optical Science and Engineering, Zhejiang University, Hangzhou 310027, China.

[#]State Key Laboratory of Precision Spectroscopy, East China Normal University, Shanghai 200062, China.

[II]Materials and Structures Laboratory, Tokyo Institute of Technology, 4259 Nagatsuta, Midori-ku, Yokohama 226-8503, Japan.

[*]Correspondence should be addressed to X.L. (xfliu@zju.edu.cn) and J.Q. (qjr@zju.edu.cn).

Epsilon-near-zero (ENZ) materials, with the relatively dielectric permittivity that is close to zero, have recently drawn broad attention from photonics communities for its potential in a plethora of exciting applications such as optical cloaking, subwavelength imaging and nonlinear optics.[1-12] Due to the unique dielectric condition, electromagnetic fields can be confined to subdiffractional dimension, therefore providing an available high local field that is favorable for boosting nonlinear optical processes.[5-7,9,10,13] Enhanced nonlinear optical effects in ENZ region have also been confirmed and exploited for designing novel devices and functionalities.[4,8,11,12,14-18] Transparent conducting oxides (TCOs), one kind of ENZ materials achieved from degenerately doping of wide-bandgap oxide semiconductors, have lately emerged as better material solutions to nanophotonics for its highly tunable optical properties and potentially both electrically and optically dynamical controllabilities as well as lower loss and CMOS-compatible device designing compared with noble metals.[10,14-17,19-23] Besides, TCOs can support large doping levels (carrier density as large as $\sim 10^{21}$ cm$^{-3}$), enabling the ENZ region to localize in the near-infrared bands which is highly desirable for optical modulation at technologically important telecom wavelengths.[14,21-25]

Here, we report ultrafast pulse generation by exploiting the ultrafast nonlinear transient optical response in ENZ TCO colloidal nanocrystals. Due to the unique highly confined nanostructured geometry compared with thin film, the nonlinear transient response of our TCO nanocrystals is further enhanced by the rich surface trap states, resulting in a shorter recovery time of ~450 fs corresponding a response speed of over 2 THz. Besides, the colloidal chemistry-derived routine also provides a scalable and wide-adaptive solution for printable photonics and optoelectronics. As a proof-of-concept demonstration, a mode-locked fiber laser at the telecom band is constructed based on these ENZ NCs to tune the continuous laser waves into femtosecond laser pulses.

Indium tin oxide (ITO), a most widely used transparent electrode material, have just been surprisingly discovered to bear a large nonlinearity in the ENZ region and therefore arousing strong renewed interests for photonics.[10,15,16] Here, ITO nanocrystals (NCs) were adopted as an example also for its easy tunability through doping. ITO NCs with various doping levels were synthesized through facile hot injection method (see method for details).[26] As seen from the transmission electron micrographs (TEMs) in Figure 1a, monodispersed NCs with an average diameter of ~8-9 nm (Figure 1d) is obtained for doped ITO NCs, and the crystallinity was confirmed by high resolution transmission electron micrographs (HRTEMs) and electron diffraction patterns (Figure 1b,c) (structural characterizations of undoped and other level doped NCs can also be founded in Supporting Information). It is noteworthy that the crystal structure maintains stable even with the doping level increased to 12% (see the powder X-ray diffraction pattern in Supporting Information), indicating a high solid-solubility limit of oxide semiconductors which offers a broader freedom for tuning their optical properties through doping.

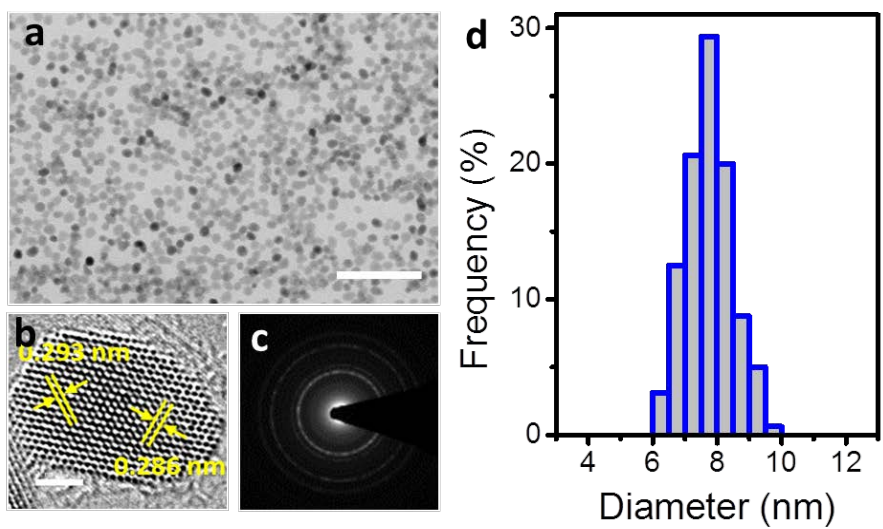

**Figure 1 Synthesis of ITO nanocrystals.** Typical transmission electron micrograph (TEM) (a), high resolution transmission electron micrograph (HRTEM) (b) and electron diffraction pattern (c) of synthesized doped ITO nanocrystals (12% Sn-doped sample ITO-12). Characterizations of other level-doped nanocrystals can be found in Supporting Information Figure S1. Scale bar in (a) and (b) are 50 nm and 2 nm, respectively. (c) The size distribution histogram of (a).

To determine the static optical response and their doping-induced tunability of these synthesized nanocrystals, the extinction spectra at various doping levels were measured and shown in Figure 2a. Clearly, doping the $In_2O_3$ NCs induces an additional infrared absorption which originates from the doping-induced electrons oscillating with incident light. And the absorption peak shows blueshift with increasing doping level, which is due to the increased electron density as a result of heavier doping according to the Drude moel (Figure 2b, see Supporting Information for more details). On the other hand, the permittivity of the ITO NC film at various doping level is presented in Figure 2c, based upon fitting of the room-temperature extinction spectra with Drude-Lorentz model (see Supporting Information for details). Benefited from the doping-induced high carrier density, the ENZ can be easily achieved in the near-infrared wavelength range. With increasing doping level and thus higher carrier density, the ENZ wavelength shifts from ~1600 nm to ~1300 nm (Figure 2d), offering a broad available tuning range in the telecom band through simply doping control, which is highly preferable for nanofabrication.

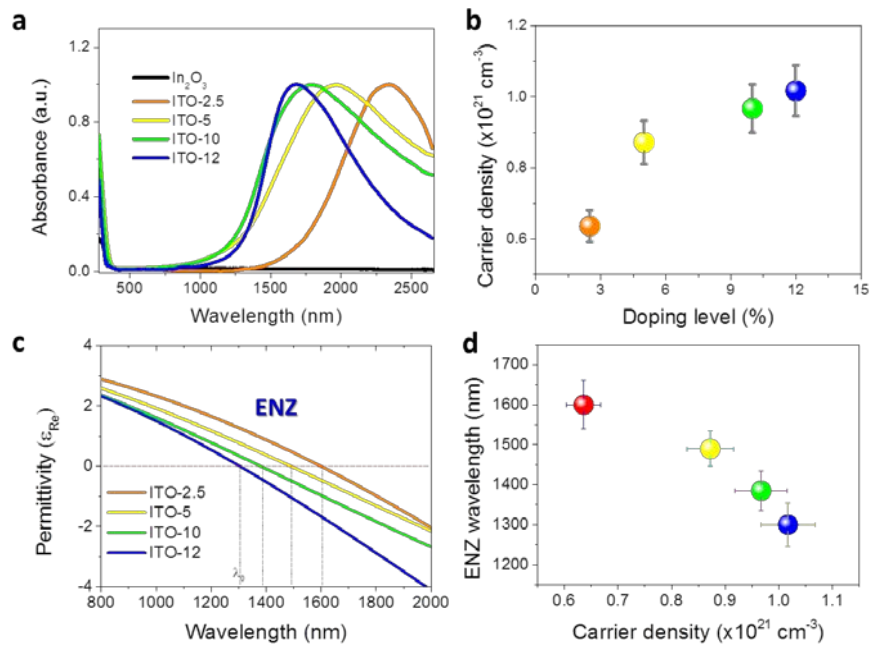

**Figure 2 Linear optical response of ITO nanocrystals.** (a) Normalized optical extinction spectra of spin-coated ITO NC thin film with various doping levels. (b) Doping level dependent carrier density. (c) Wavelength dependent relatively permittivity of the spin-coated ITO NC thin films. (d) The relationship between ENZ wavelength and carrier density.

In order to investigate the transient optical response of the synthesized ENZ NCs, we adopted a degenerate pump-probe technique to record the ultrafast dynamics of excited carriers under laser pulse excitation at the ENZ region. As an example shown in Figure 3a, upon a moderate femtosecond pulse excitation with power of as low as ~0.7 mJ/cm$^2$, a bleach with a response amplitude as large as ~22% occurs, of which, however, the same modulating level could be only achieved with excitation power as large as ~7 mJ/cm$^2$ in ITO nanorod arrays[16]. Considering that the typical damage threshold for oxide dielectrics are at the level of ~100 mJ/cm$^2$ for femtosecond pulses, this effect could be further enhanced, indicating a superior nonlinear optical performance of our highly confined nanostructured ITO NCs. Besides, the transient effect can be fully recovered within ~450 fs, corresponding to a modulation speed over 2 THz, which is faster than recent reported Al-doped ZnO (AZO) film and ITO film/nanorod arrays.[14-16] The faster dynamics shall be also a result of the unique nanosized highly confined geometry that is intrinsic of rich surface multiple trapping states.[27,28]

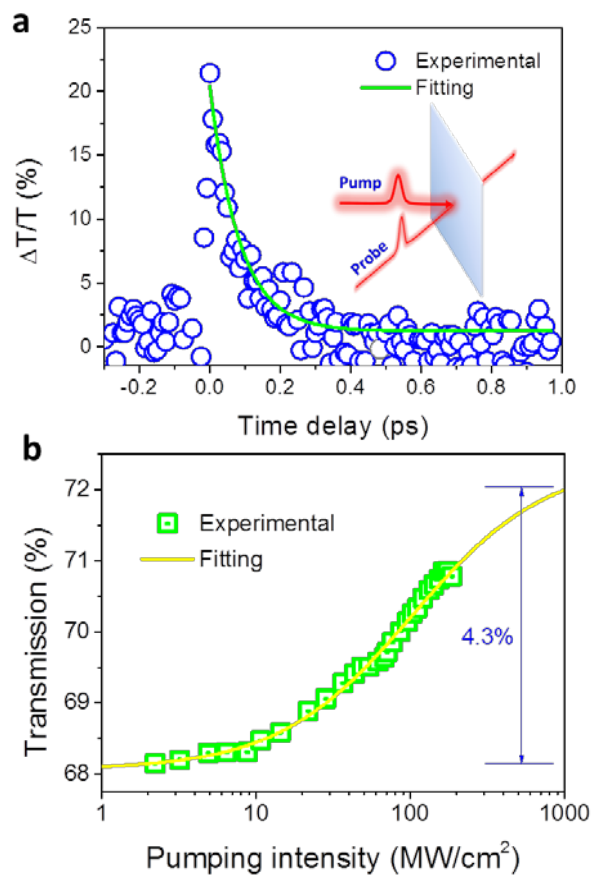

**Figure 3 Nonlinear optical response of ITO nanocrystals at ENZ region**. (a) Transient bleach

dynamics of ITO NCs. Solid line shows a close fit with a single exponential decay behavior, of which the fitted decay lifetime is ~ 84 fs. Importantly, the sub-100 fs transient with negligible signal at longer times promises a high-fidelity modulation at terahertz frequencies. Inset schematically shows the optical setup. The pump and the probe wavelengths are both at 1.3 μm near ENZ wavelength with a pulse width of ~35 fs. (b) Input dependent transmission of a constructed ITO/PMMA composite film measured with a pulsed laser at 1.55 μm with pulse width of ~500 fs. The data can be fitted with a saturable absorption curve as $1 - \mathrm{T} = \alpha_s/(1 + I/I_s) + \alpha_{ns}$, where $\alpha_s = 4.3\%$, $\alpha_{ns} = 27.6\%$, $I_s = 103$ MW/cm$^2$ are the modulation depth, non-saturable loss, and saturable absorption intensity, respectively.[29,30]

The ultrafast transient bleach dynamics and high efficient modulation under laser pulse excitation lead us to explore the possibility of all-optical modulating applications with these NCs. For the ease of device integration, nanosized materials are usually dispersed in a polymer matrix to form a composite thin film which affords great freedom and flexibilities for device design and fabrication. Considering both the high transparency and thermal and chemical stabilities, poly(methyl methacrylate) (PMMA) was chosen as the matrix for ITO NCs.[31] Obviously, as shown in Figure 3b, the transmission of this thin film increases with the input laser intensity, which can be fitted with a saturable absorption behavior (Figure 3b). It should be noted that the modulation depth, non-saturable loss are related to the concentration of NCs in PMMA, which, in other words, could be further optimized with proper controlling of the ratio of ITO NCs to PMMA. Besides, the pulse properties also have an important effect on the modulation depth, where the longer pulse width may results in smaller modulation depth but should be closer to the case in

practical applications such as pulse generation.[31]

We then inserted the constructed ITO/PMMA film into a fiber laser to build a saturable absorber, as shown in Supporting Information. Under a pumping of ~6 mW, continuous wave was achieved at ~1560 nm (Figure 4a). Upon increasing the pumping power to ~18 mW, the intensity of the propagation light in the waveguide is large enough for the ITO film to trigger a mode-locking operation, which results in squeezing the continuous wave into laser pulses due to the so-called saturable absorption. Amazingly, a stable (with signal-to-noise of ~56 dB, Figure 4c) femtosecond pulse train with temporal width as short as ~593 fs was obtained by taking out the pulses with a 10% coupler and identified with a high-speed autocorrelator. It is noted that shorter pulses could be expected by further dispersion management of the waveguide structure. Importantly, as molecular motions and chemical reactions with electron and energy transfer occurs in the femtosecond to picosecond regime, laser pulses with subpicosecond temporal width could be used as a probe to even tracking the motions of a single molecular or electron.[32-34] On the other hand, shorter pulses also bring new opportunities for communication technologies as well as medical imaging, precision metrology, micromachining, etc.[35,36]

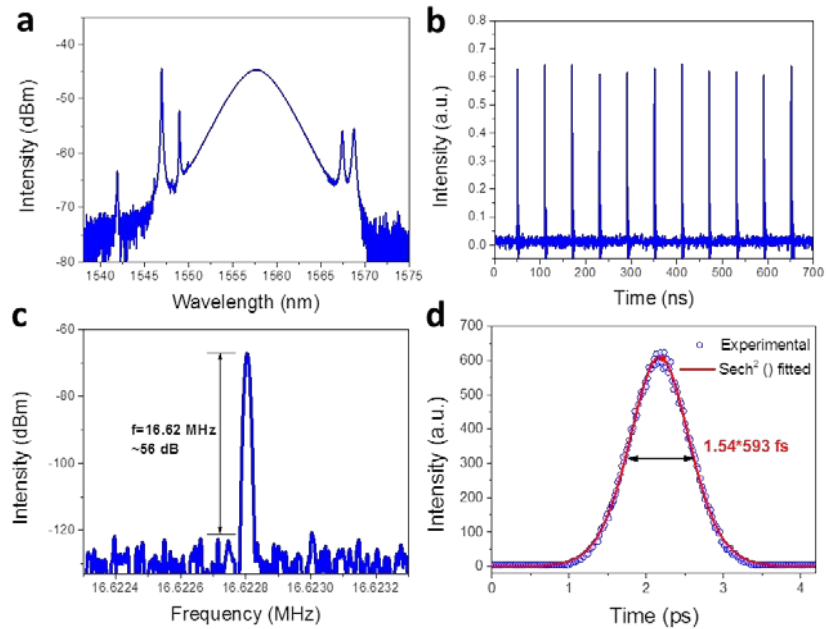

**Figure 4 Ultrafast pulse generation with ITO nanocrystals at telecom band.** (a) Optical spectrum. (b) Mode-locking pulse train. (d) Autocorrelation trace. (d) RF optical spectrum.

In conclusion, we have shown that colloidal ITO nanocrystals with ENZ wavelength tunably across the telecom band enabled by controllable doping are excellent candidates for ultrafast pulse generation. The highly confined nanostructured geometry compared with thin films provides an intrinsic advantage in enhancing the ultrafast nonlinear processes in terms of both response amplitude and speed. An ultrafast optical fiber modulator based on these ENZ NCs was constructed to tune continuous laser wave into femtosecond laser pulses. Besides, the colloidal chemistry-derived routine also provides a scalable and wide-adaptive solution for printable and flexible photonics and optoelectronics. Therefore, our colloidal ENZ NCs may serves as a promising material solution for realizing advanced dynamic devices.